\documentclass[12pt]{article}
 \usepackage{amsmath}
 \usepackage{amsfonts}
 \usepackage{amssymb}
 
 \usepackage{pst-plot}
 \usepackage{pstricks}
 \usepackage{multido}
 \usepackage{fp}
 \usepackage{ifthen}
 \usepackage{graphicx}
 \usepackage{subfig}

 \usepackage{url}

 \allowdisplaybreaks[4]     

 \renewcommand{\L}{Lema\^{\i}tre}
 \newcommand{\LT}{\L-Tolman}

 \newcommand{\showlabel}[1]{
   \label{#1}
 }
 
 \newcommand{\Ra}{\Rightarrow}

 \newcommand{\nn}{\nonumber}
 \newcommand{\er}[1]{(\ref{#1})}          

 \newcommand{\bs}[1]{\mbox{\boldmath $#1$}}

 \newcommand{\Rt}{\dot{R}}

 \renewcommand{\d}{{\rm d}}     
 \newcommand{\p}{\partial}     
 \newcommand{\td}[2]{\frac{{\rm d} #1}{{\rm d} #2}}     
 \newcommand{\pdil}[2]{\partial #1/\partial #2 }     

 \newcommand{\WW}{E'/E}

\begin{document}
 \sffamily

 \title{Symmetry and Equivalence in Szekeres Models}
 
 \author{
   Ira Georg
   \thanks{\tt GRGIRA001@myuct.ac.za} \\
   \and
   Charles Hellaby
   \thanks{\tt Charles.Hellaby@uct.ac.za} \\
   {\small \it Dept. of Maths. and Applied Maths,
   University of Cape Town,
   Rondebosch,
   7701,
   South Africa}
 }
 
 \date{}

 
 \maketitle

 \begin{abstract}
 We solve for all Szekeres metrics that have a single Killing vector.  For quasi hyperboloidal ($\epsilon = -1$) metrics, we find that translational symmetries are possible, but only in metrics that have shell crossings somewhere, while metrics that can be made free of shell crossings only permit rotations.  The quasi planar metrics ($\epsilon = 0$) either have no Killing vectors or they admit full planar symmetry.  Single symmetries in quasi spherical metrics ($\epsilon = +1$) are all rotations.  The rotations correspond to a known family of axially symmetric metrics, which for each $\epsilon$ value, are equivalent to each other.  We consider Szekeres metrics in which the line of dipole extrema is required to be geodesic in the 3-space, and show the same set of families emerges.  We investigate when two Szekeres metrics are physically equivalent, and complete a previous list of transformations of the arbitrary functions.
 \end{abstract}

 \section{Motivation}

  In 1975 Peter Szekeres \cite{Szek75a,Szek75b} discovered a very interesting family of exact inhomogeneous solutions of the Einstein field equations, for which the matter source is a comoving, zero pressure fluid (dust).  There are 6 arbitrary functions that depend on the ``radial" coordinate%
 \footnote{Here the ``radial" coordinate is the one that becomes a true coordinate radius in the spherically symmetric special case.}%
 .  Although one of these functions can be used to remove the re-scaling freedom in the ``radial" coordinate, there is no canonical choice that does not restrict the physical possibilities \cite{Hel09}.  

 There are actually two classes of Szekeres model, the more commonly used one is a generalisation of the \LT\ \cite{Lem33,Tol34} and Ellis \cite{Elli67} metrics, and the other is a generalisation of the Datt-Kantowski-Sachs \cite{Datt38,KanSac66} metrics.  Since the latter can be viewed as a limit of the former \cite{Hell96b}, we focus on the \LT-Ellis-like metric.  Three of the Szekeres arbitrary functions are identical to those of the ``underlying" \LT or Ellis metric, and the other three control the deviation from spherical, planar, or hyperboloidal (pseudo-spherical) symmetry.

 A key result for this metric was the proof by Bonnor, Suleiman and Tomimura (BST) that it has no Killing vectors \cite{BoSuTo77}.  This is despite the constant time 3-surfaces being conformally flat \cite{BeEaOl77}, and the lack of gravitational radiation \cite{Bonn76b}.  In fact the BST paper only considered the quasi-spherical case, though one would expect it to generalise; a task which we complete here along the way.

 Nolan and Debnath \cite{NolDeb07}, in investigating shell focussing singularities, have shown that if a quasi-spherical Szekeres spacetime has a ``radial" null geodesic, then the spacetime is axially symmetric and the ray lies along the axis; if there is more than one radial null geodesic, it is spherically symmetric.  Further, they showed that all axi-symmetric Szekeres models are equivalent.

 Krasinski \& Bolejko \cite{KraBol11} considered light paths and redshifts in Szekeres models.  In general, light rays emitted from the same matter point at different times, and received by the same observer, do not follow the same comoving spatial path.  The authors asked under what conditions the spatial paths of two such light rays might be repeated.  They showed that if all light paths between every emitter-observer pair are repeated the spacetime must be a (dust) FLRW model, and if there exists a repeated ray along a single direction then the model is axi-symmetric about that direction.  They also generalised the Nolan and Debnath result to quasi-hyperbolic and quasi-planar models.  For the \LT\ and Ellis models, only the ``radial" rays have repeatable paths.

 Neither of the above papers claimed they had found all the axi-symmetric Szekeres models, and the possibility, in the quasi-hyperbolic and quasi-planar models, of other, non-rotational single symmetries was not considered.

 Sussman \& Gaspar \cite{SusGas15} studied the location of extrema of density, expansion and spatial curvature, and they produced some very nice numerical examples and plots of Szekeres models with complex structure.  They also mentioned cases where one or two of the non-spherical arbitrary functions are constant, and suggest some are axially symmetric.  More complex Szekeres matter distributions --- networks of matter structure --- were investigated by Sussman, Gaspar \& Hidalgo \cite{SuGaHi15}.

 Though there is a significant body of work on the Szekeres models,
 \cite{Apos16a,Apos16b,ApoCar07,BarSte84,BeEaOl77,Bole06a,Bole07,Bole08a,Bole08c,Bole10,BolCel10,BolNazWil16,BolSus11,BolWyi08,Bonn76b,Bonn86,BonPug87,BoSuTo77,BonTom76,ChaDeb08,Cova80,Debn11,deS85,Glei84,GooWai82a,GooWai82b,Hell96b,HelKra02,HelKra08,HelWal12,HePrIbCa12,IshPee12,IRGWNS08,JosKro96,KokHan15,Kra08,Kra16,KraBol11,KraBol12a,KraBol12b,MeuBru11,MisCel14,MiCeSi12,NolDeb07,NwIsTh11,PeIsTr12,PeTrIs14,SusBol12,SusGas15,SuGaHi15,Szek75a,Szek75b,Vill14,VrbSvi14,Wain77,WalHel12},
 and there has been a renewal of interest recently, the lack of Killing vectors makes this metric relatively hard to work with.  The quasi-planar and quasi-hyperboloidal cases have been especially neglected.  Therefore the symmetric special cases with a single Killing vector would be useful as stepping stones between full spherical, planar or hyperboloidal symmetry and the general Szekeres case.

 A fuller description of the Szekeres metric and its properties can be found in \cite{Kra97,Hel09,PleKra06}.

 \section{The Szekeres Metric}

 The line element is
 \begin{align}
   \d s^2 = - \d t^2 + \frac{(R' - R E'/E)^2}{(\epsilon + f)} \d r^2 + \frac{R^2}{E^2} \Big( \d p^2 + \d q^2 \Big) ~,
   \showlabel{ds2Sz}
 \end{align}
 where $\epsilon = -1, 0, +1$, $f = f(r)$, $E = E(r, p, q)$, $R = R(t,r)$ and we write
 \begin{align}
 W = \sqrt{\epsilon + f}\; ~.  
 \end{align}
 The evolution function $R$ obeys a Friedmann equation
 \begin{align}
   \Rt^2 = \frac{2 M}{R} + f + \frac{\Lambda R^2}{3} ~,   \showlabel{RtSqEq}
 \end{align}
 where $M = M(r)$, and solving this DE introduces the bang-time function $t_B = t_B(r)$.  
 The matter is comoving, $u^a = \delta^a_t$, and has a dust equation of state ($p = 0$), the density being given by
 \begin{align}
   \kappa \rho & = \frac{2 (M' - 3 M E'/E)}{R^2 (R' - R E'/E)} ~.
   \showlabel{MassDensity}
 \end{align}
 Each 2-surface of constant $r,t$,
 \begin{align}
  dl^2= \frac{(\d p^2 + \d q^2)}{E^2} ,  \showlabel{ds2Unit}
 \end{align}
 is a unit 2-sphere if $\epsilon = +1$, a unit 2-pseudo-sphere (or hyperboloid) if $\epsilon = -1$, and a 2-plane if $\epsilon = 0$.  The shape function $E$ can be written as
 \begin{align}
   E = \frac{S}{2} \left\{ \frac{(p - P)^2}{S^2} + \frac{(q - Q)^2}{S^2} + \epsilon \right\} ~,   \showlabel{Edef}
 \end{align}
 where $S = S(r)$, $P = P(r)$, $Q = Q(r)$, and $(p, q)$ are stereographic coordinates on each 2-surface.
 The transformation between stereographic coordinates $(p,q)$ and the more regular ``polar" type coordinates can be found in \cite{HelKra08} for each of the three $\epsilon$ values.  

 $R(t,r)$ gives the evolving scale area of the constant $r$ shell, since its square multiplies the unit-scale surface, \er{ds2Unit}.  We will refer to $r$ as a ``coordinate radius" and $R$ as an ``areal radius", for all $\epsilon$ values, it being understood that the pseudo-spherical or planar equivalents are intended if $\epsilon \neq +1$.

 The function $f(r)$ determines the curvature of the constant $t$ 3-spaces, and it also gives the expansion energy per unit mass, ${\cal E} = f/2$, of the particles at ``radius" $r$.  For $\epsilon = +1$, $M(r)$ is the total gravitational mass interior to the comoving shell $r$; for other $\epsilon$ values, it is a mass-like factor in the gravitational potential energy.  The functions $S(r)$, $P(r)$ \& $Q(r)$ determine the strength and orientation of the dipole on each comoving shell.

 The function $S$ cannot be zero \cite{HelKra02,HelKra08}, so it is convenient to keep it positive, $S > 0$.  If $\epsilon = 0$ globally, then $S$ may be absorbed into the other arbitrary functions, so one is free to set $S = 1$.  In order to avoid shell crossing singularities \cite{HelKra02}, the arbitrary functions must obey a set of conditions that limit their ranges relative to each other; these are not very restrictive.  

 Both $g_{rr}$ in \er{ds2Sz} and $\rho$ in \er{MassDensity} depend on $\WW$, which varies over each 2-surface of constant $(t, r)$, and has extreme values
 \begin{align}
   \left. \frac{E'}{E} \right|_e = \pm \frac{\sqrt{S'^2 + \epsilon( P'^2 + Q'^2)}\;}{S} ~,
      \showlabel{ErEeS}
 \end{align}
 located at
 \begin{align}
   p_e & = P + \frac{ P'}{(S'/S) + (\WW)_e} ~,   \showlabel{pe} \\
   q_e & = Q + \frac{ Q'}{(S'/S) + (\WW)_e} ~.   \showlabel{qe}
 \end{align}

 The quasi-spherical Szekeres metric, with $\epsilon = +1$, is commonly described as an assembly of evolving spheres which are non-concentric, and which display a dipole distribution in the density variation around each sphere.  The dipole is due to the factor $E'/E$; on each $(p, q)$ 2-sphere, it is zero on an ``equator", maximum at one ``pole", and minimum at the opposite ``pole", with $E'/E|_{\text{min}} = - E'/E|_{\text{max}}$.  The strength and orientation of the dipole depend on coordinate radius $r$ through \er{ErEeS}-\er{qe}.

 The quasi-hyperboloidal Szekeres metric, with $\epsilon = -1$, has been much less studied.  As shown in \cite{HelKra08}, it may be thought of as an assembly of evolving right-hyperboloids that are stacked non-symmetrically or ``non-concentrically".  The two sheets of the hyperboloid map to separate regions of the $(p, q)$ plane, one inside the boundary circle
 \begin{align}
   (p - P)^2 + (q - Q)^2 = S^2 ~,   \showlabel{BndCrc}
 \end{align}
 and one outside it.  The boundary circle is the locus of infinity for each sheet.  Only one of the two sheets can be free of shell crossings and only if
 \begin{align}
   S'^2 > P'^2 + Q'^2 ~.   \showlabel{QHSzCondit}
 \end{align}
 In this latter case too there is a kind of hyperboloidal (or pseudo-spherical) dipole with strength (\ref{ErEeS}) and orientation defined by (\ref{pe}) and (\ref{qe}).  If \er{QHSzCondit} is not satisfied, \er{ErEeS} is not real, and extrema with respect to $p$ \& $q$ do not exist.
 
 The quasi-planar Szekeres metric, given by $\epsilon = 0$, does not have an extremum of $E'/E$, so one cannot talk about a ``dipole".

 It should be noted that a single spacetime can have both $\epsilon = +1$ and $\epsilon = -1$ regions, joined by an $\epsilon = 0$ region which may be thin (a 3-surface) or may have finite width.

 \section{Szekeres Models With A Single Symmetry}
 \showlabel{SzSingSymm}

 The Szekeres metric --- in its fully general form --- has no Killing vectors \cite{BoSuTo77}.  However, it contains the spherically-symmetric special case, the \LT\ model \cite{Lem33,Tol34}. For example, if we choose the functions $S$, $P$ \& $Q$ to be constant
 \begin{align}
   \epsilon = +1 ~,~~~~~~ 0 = S' = P' = Q' ~,   \showlabel{SphSymCnd}
 \end{align}
 we will find a \LT\ model.
 Therefore we expect that axially symmetric special cases exist; and indeed examples such as $\epsilon = +1$, $P' = 0$, $Q' = 0$ are known.
 
 The spherical symmetry of LT models results from the fact that every const.\ $r,t$ submanifold is spherically symmetric with respect to a common center. Although a more general quasi-spherical ($\epsilon=1$) Szekeres model also has spherically symmetric $r,t$ constant submanifolds%
 \footnote{for short we say ``spheres" if $\epsilon =1$, ``hyperboloids" if $\epsilon =-1$, simply ``planes" if $\epsilon =0$, or ``shells" if $\epsilon $ is not specified, instead of ``$r,t=$ const.\ submanifolds"},
 their centers do not coincide.  
 When we consider an axially-symmetric arrangement of non-concentric spheres, then there is only one possibility: the centers of the spheres ought to be in a ``straight line'', a geodesic, that forms the symmetry axis. 

 In quasi-spherical models, the dipole function $\WW$ encodes the distance between two neighbouring spheres with respect to the mean distance at the equator, see figure \ref{fig::spheres}. If the dipole function is dependent on the coordinates on the sphere $p,q$, then the distance to the next sphere (with label $r+dr$) is different for different $p,q$. The two neighbouring spheres are non-concentric and because of the spherical symmetry of the two shells, $\WW$ must show a dipole structure (see figure \ref{fig::pErE}), i.e.\ $\WW$ is axially symmetric (for constant $r$), the points of extrema are antipodal, and the extremal values are equal with opposite signs.
 If $\WW$ is constant on all spheres, the model is necessarily spherically symmetric. $\WW$ can be considered as the deviation function from global spherical symmetry.

 We expect an axial symmetry to occur if the centers of the $r,t=$ const. submanifolds are displaced only along a ``straight line", i.e.\ if the extrema of $\WW$ form a geodesic.  
 Similarly, for the quasi-planar and quasi-hyperboloidal cases, we expect sub-models that are symmetric, if the dipole functions of different shells align.

 \begin{figure}
  \centering
  \subfloat[\showlabel{fig::spheres}  Two const.\ $r,t$ submanifolds in a general quasi-spherical Szekeres model. The indicator {\it min,max} mark the points of extremal distance between the spheres. The dashed curves on the inner sphere are curves of latitude with constant $\WW$. ]{
   \begin{pspicture}(-3,0)(5,4)
   \pscircle(2,2){2}
   \pscircle(1.5,2){1}
   \psellipticarc[linestyle=dashed]{-}(1.9,2)(0.4,0.93){-75}{75}
   \psellipticarc[linestyle=dashed]{-}(1.5,2)(0.3,1){-90}{90}
   \psellipticarc[linestyle=dashed]{-}(0.9,2)(0.4,1){-80}{80}
   \psellipticarc[linestyle=dashed]{-}(0.6,2)(0.2,0.5){-80}{80}
   \psline(0,2)(0.5,2)
   \rput(0.3,1.9){\tiny \it min}
   \psline(2.5,2)(4,2)
   \rput(2.7,1.9){\tiny \it max}
  \end{pspicture}
  }
  \qquad
  \subfloat[\showlabel{fig::pErE}Contour lines of $\WW$ after a stereographic projection relative to fig \ref{fig::spheres} onto the $p,q$-plane for an arbitrary choice of parameters.
 ]{
    \makebox[7cm][c]{
      \includegraphics[height=4cm]{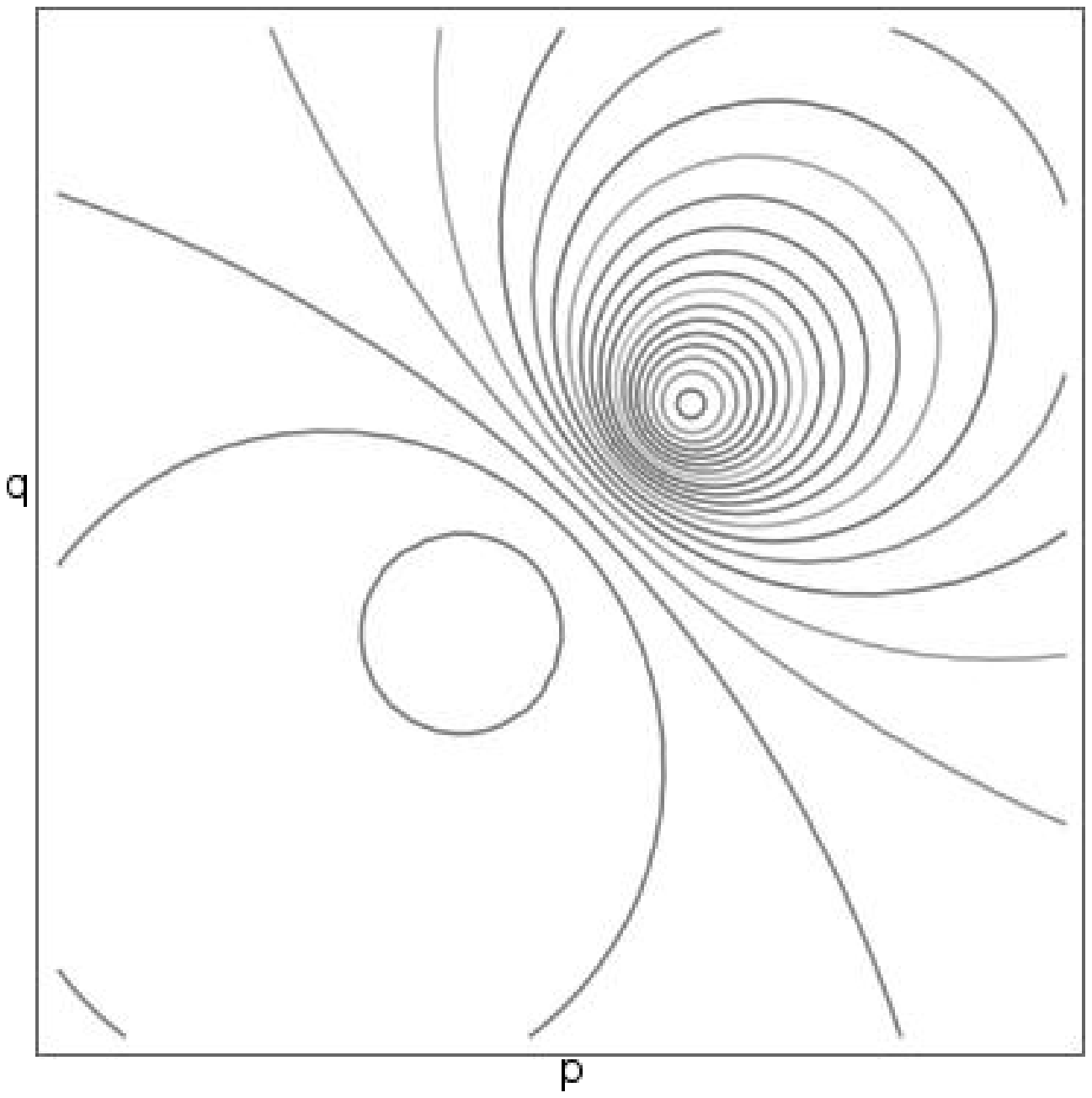}
    }
  }
  \caption{Form of $\WW$ on the sphere (a) and on the $p,q$ plane (b).
  }
\end{figure}

 In the following we will represent shells of constant $r,t$ as $p,q$-planes in a Cartesian-like plot and thus unify the representation of the three different types of models ($\epsilon=1,0,-1$). 
 Below we solve for the general case, rigorously deriving the conditions for an axial or other single symmetry from the Killing equations. 

 \subsection{The Killing Equations}
 \showlabel{KlEqs}

 We now want to formalize the above and assume that there is one Killing vector field $\xi^\mu$, i.e.\ that the metric is invariant under the infinitesimal transformation $x^\mu \rightarrow x^\mu+ \kappa \xi^\mu$ ($\kappa$ is an infinitesimal quantity). Then $\xi^\mu$ satisfies
 \begin{align}
  0 = \xi_{\nu;\mu} + \xi_{\mu;\nu} ~.
  \showlabel{KlngVcDf}
 \end{align}
 Demanding the existence of one Killing vector field will lead to conditions on three of the free functions. We are only interested in the three free functions $S,P,Q$, because the other two are already studied in the LT literature. We assume that
 \footnote{BST assumed $W$ and $M$ are linearly independent, which suffices for $\epsilon = +1$, but not in general.  The relation $f \propto M^{3/2}$ plus $t_B =$~constant is the FLRW case.}
 \begin{align}
   f,M ~\mbox{and}~ \frac{f}{M^{2/3}} ~\mbox{are not constant} ~,   \showlabel{fnotM32}
 \end{align}
 and also
 \begin{align}
   & \mbox{$S,P,Q$ do not depend on $M,f,t_B$} ~.   \showlabel{SPQIndepMftB}
 \end{align}

 We follow the calculation given in Bonnor, Suleiman \& Tomimura (BST) \cite{BoSuTo77} with two exceptions.
 First, we want to keep $\epsilon$ general, whereas they specified $\epsilon = +1$. Second, we insist that exactly one Killing vector field exists, whereas \cite{BoSuTo77} show that for general $S,P,Q$ there is no Killing vector and thus no symmetry.  Adopting the notation of BST, let us write the Szekeres metric \er{ds2Sz} in the general form
 \begin{align}
   \d s^2 & = - \d t^2 + e^\lambda \, \d r^2 + e^\omega \, (\d p^2 + \d q^2)   \showlabel{ds2SzGen}
 \end{align}
 where $\lambda = \lambda(t, r, p, q)$, $\omega = \omega(t, r, p, q)$.  More explicit forms of the metric functions will be used after the equations have been simplified.

 Let us now assume a non-vanishing Killing vector field of the form
 \begin{align}
   \xi^a = (\delta, \alpha, \beta, \gamma) ~,   \showlabel{KVform}
 \end{align}
 where each component is an unspecified function of all four coordinates.  Appendix \ref{BSTKV} collects the relevant parts of the argumentation in \cite{BoSuTo77}, generalising it to all $\epsilon$, and here we just note the key results.  If we assume the LT functions $M$ and $f$ are arbitrary, we are left with a Killing vector of the form
 \begin{align}
   \xi^a = \big( 0, 0, \beta(p, q), \gamma(p, q) \big) ~.   \showlabel{BtGmpq}
 \end{align}
 It is remarkable that a Killing vector field in a Szekeres model can have no $r$-component, and its two non-vanishing components must be independent of $r$. In these coordinates the Killing vector field looks the same for all $r,t$. Now, let us find the Killing vector field and thus the restriction on the functions $S,P,Q$.
 
 In BST's derivation of (\ref{BtGmpq}), four of the Killing equations were not fully used and still have information that we can further exploit. Let us rewrite the remaining Killing equations
 \begin{align}
   & rr\mbox{-component} & \beta \, \lambda_{,2} + \gamma \, \lambda_{,3} & = 0   \showlabel{Kerr} ~, \\
   & pp\mbox{-component} & \beta_{,2} + \frac{1}{2} \big( \beta \, \omega_{,2} + \gamma \, \omega_{,3} \big) & = 0
      \showlabel{Kepp} ~, \\
   & qq\mbox{-component} & \gamma_{,3} + \frac{1}{2} \big( \beta \, \omega_{,2} + \gamma \, \omega_{,3} \big) & = 0
      \showlabel{Keqq} ~, \\
   & pq\mbox{-component} & \beta_{,3} + \gamma_{,2} & = 0 ~,   \showlabel{Kepq}
 \end{align}
 where $\lambda_{,2} = \pdil{\lambda}{p}$, $\omega_{,3} = \pdil{\omega}{q}$, etc.
 It can be seen from \er{Kepq} and the result of subtracting \er{Keqq} from \er{Kepp},
 \begin{align}
   \beta_{,2} - \gamma_{,3} & = 0 ~,   \showlabel{Keppqq}
 \end{align}
 that $\beta$ and $\gamma$ are conjugate harmonic functions.

 We know that if $\lambda_{,2} = 0$ and $\lambda_{,3} = 0$ we have a global spherical ($\epsilon=1$), planar ($\epsilon=0$) or pseudo-spherical ($\epsilon=-1$) symmetry, since the $rr$ component of the metric loses its dependency on the coordinates $p,q$. The geometry of the 2 dimensional submanifolds is passed on to the 3 dimensional space.
 We cannot have only one of the two $\lambda_2,\lambda_3$ vanishing everywhere, because it would result in $0 = \epsilon S' = P' = Q'$, hence $\WW=0$, and so the other of the two would also vanish.

 We now assume that both $\lambda_2$ and $\lambda_3$ are non-zero.%
 \footnote{When we say that a function is non-zero, we mean that it does not vanish everywhere, although there might be a submanifold where this function vanishes.}
 Then equation \er{Kerr} gives us a way to obtain $\gamma$ from $\beta$, and convert \er{Kepp} into an exponential differential equation in $\beta$ alone,
 \begin{align}
   0 & = \beta_{,2} + \frac{1}{2} \left( \omega_{,2} + \frac{\lambda_{,2}}{\lambda_{,3}} \, \omega_{,3} \right) \beta ~.
      \showlabel{beta2Eq}
 \end{align}
 In order to derive the Killing vector explicitly, we use the abbreviations
 \begin{align}
   Y = p - P ~,~~~~~~ Z = q - Q ~,~~~~~~ A = 2 S E = Y^2 + Z^2 + \epsilon S^2 ~,~~~~~~ \beta = e^H ~,
 \end{align}
 and write the solution as
 \begin{align}
   H & = \int \frac{\p_2 \beta}{\beta} \, \d p = - \frac{1}{2} \int \left( \omega_{,2}
      + \frac{\lambda_{,2}}{\lambda_{,3}} \, \omega_{,3} \right) \, \d p \\
   & = - 2 \int \left( \frac{Y \, Z' - Y' \, Z}{Z \, A' - Z' \, A} \right) \d p
      = \int \frac{\p_2 \big( Z \, A' - Z' \, A \big)}{\big( Z \, A' - Z' \, A \big)} \d p \\
   \Ra~~~~~~~~ \beta & = h \big( Z \, A' - Z' \, A \big) ~,   \showlabel{BtSln0}
 \end{align}
 where $h(r,q)$ is a function of integration independent of $p$.
 Similarly \er{Kerr} in \er{Keqq} gives
 \begin{align}
   0 & = \gamma_{,3} + \frac{1}{2} \left( \frac{\lambda_{,3}}{\lambda_{,2}} \, \omega_{,2} + \omega_{,3} \right) \gamma \\
   \Ra~~~~~~~~ \gamma & = g \big( Y \, A' - Y' \, A \big) ~,   \showlabel{GmSln0}
 \end{align}
 where $g(r,p)$ is a function of integration independent of $q$.  Putting \er{BtSln0} \& \er{GmSln0} in \er{Kerr} we find
 \begin{align}
   -1 & = \frac{\beta}{\gamma} \, \frac{\lambda_{,2}}{\lambda_{,3}}
   = \frac{h(r,q)}{g(r, p)}
 \end{align}
 so clearly
 \begin{align}
   h = h(r) = - g(r) ~.
 \end{align}
 We find the Killing vector components to be
 \begin{align}
   \beta & = h \Big[ 2 \, Z \big( - Y \, P' - Z \, Q' + \epsilon \, S \, S' \big)
      + Q' \big( Y^2 + Z^2 + \epsilon \, S^2 \big) \Big]   \showlabel{BtSln1} \\
     & = h \Big[ Q' \, (p^2 - q^2) - 2 \, P' \, p \, q + 2 (Q \, P' - P \, Q') p + 2 (P \, P' + Q \, Q'
         + \epsilon \, S \, S') q \nn \\
     &~~~~~~~ - 2 \, P \, Q \, P' + (P^2 - Q^2 + \epsilon \, S^2) Q' - 2 \, Q \, \epsilon \, S \, S' \Big]
        \showlabel{BtSln2} \\
   \gamma & = - h \Big[ 2 \, Y \big( - Z \, Q' - Y \, P' + \epsilon \, S \, S' \big)
      + P' \big( Y^2 + Z^2 + \epsilon \, S^2 \big) \Big]   \showlabel{GmSln1} \\
     & = - h \Big[ P' \, (q^2 - p^2) - 2 \, Q' \, p \, q + 2 (P \, Q' - Q \, P') q + 2 (P \, P' + Q \, Q'
      + \epsilon \, S \, S') p \nn \\
      &~~~~~~~~~ - 2 \, P \, Q \, Q' + (Q^2 - P^2 + \epsilon \, S^2) P' - 2 \, P \, \epsilon \, S \, S' \Big] ~.
        \showlabel{GmSln2}
 \end{align}
 The function $h$ is constrained by the fact that the Killing vector components are independent of $r$ as required by (\ref{BtGmpq}): $\beta' = 0 = \gamma'$. Therfore we must have the coefficients of the different powers of $p$ \& $q$ in $\beta'$ \& $\gamma'$ vanish.
 A short calculation gives us the following three conditions:
 \begin{align}
	h'P'+hP''&= 0 ~, \\
	h'Q'+hQ''&= 0 ~, \\
	2h' \epsilon SS'+2h \epsilon SS'' +2h(P'^2+Q'^2+ \epsilon S'^2)&= 0  ~.
 \end{align}
 After integrating, we find
 \begin{align}
	hP'&= c_p \showlabel{KVcond1} ~, \\
	hQ'&= c_q \showlabel{KVcond2} ~, \\
	h (\epsilon SS' +PP' +QQ') &= c_s ~, \showlabel{KVcond3}
 \end{align}
 where $c_p,c_q,c_s$ are constants of integration. Also note that $h\neq 0$ or we would have vanishing Killing vector components (\ref{BtSln1}) and (\ref{GmSln1}).
 
 From these we get
 \begin{align}
   P = c_p \int \frac{\d r}{h} + c_{p0} ~,~~~~ Q = c_q \int \frac{\d r}{h} + c_{q0} ~,~~~~
   \epsilon S^2 = - P^2 - Q^2 + 2 \, c_s \int \frac{\d r}{h} + c_{s0} ~,   \showlabel{PQSint}
 \end{align}
 which shows there is only one free function between $S$, $P$ and $Q$.

 So far we have found the Killing vector field for a Szekeres model, equations \er{BtSln0} and \er{GmSln0}, and three conditions that constrain $h$, $S$, $P$, and $Q$, \er{KVcond1}-\er{KVcond3}.
 Let us be more explicit and consider 3 different cases: 
 \paragraph{\bf Case 1:}~~ $P'=0=Q'$.~~ The first two conditions, (\ref{KVcond1}) and (\ref{KVcond2}), are trivial. The third, (\ref{KVcond3}), can be inserted in $\beta,\gamma$ for $\epsilon=\pm 1$.  There is no further constraint on $S$. If $\epsilon=0$, $P'=Q'=0$ leads to $\lambda_{,2}=\lambda_{,3}=0$, which we excluded as the full symmetric model.
 \paragraph{Case 2:}~~  Let one of $P',Q'$ be zero and the other one nonzero, say $P'\neq 0$, $Q'= 0$.~~ Then we can use the first condition and find $h=c_p/P'$. The third condition then gives $\epsilon SS' =  - P P'+(c_s/c_p) P'$ and after integrating
 \begin{align}
   \epsilon S^2 &= -P^2 +2 c_2 P + c_3 ~.
   \showlabel{Case2Cond}
 \end{align}
 If $\epsilon=0$ we find $0=P'(c-P)$ and we must have $P'=0$ contrary to the assumption of case 2. There is no case 2 Killing vector for $\epsilon=0$.

 Equivalently if $Q' \neq0$, $P'= 0$ we find $\epsilon S^2 = -Q^2 +2 c_2 Q + c_3$ and no Killing vector for $\epsilon =0$.
 \paragraph{Case 3:}~~ Let both $P',Q'\neq 0$.~~ Then the first two conditions give us $h=\frac{c_q}{Q'} =\frac{c_p}{P'}$ and therefore $Q'=cP'$ with $c=\frac{c_q}{c_p}$. The second and third conditions then become, after renaming the constants,
 \begin{align}
    Q = c P + c_Q ~,~~~~~~
    \epsilon S^2 &= -(1+c^2)P^2 +2 c_2 P + c_3 ~.
    \showlabel{Case3Cond}
 \end{align}
 If $\epsilon=0$ we find from the first two conditons $Q'=\frac{c_q}{c_p}P'$ and the third leads to $Q'=-\frac{c_p}{c_q}P' $ and thus $c_p=c_q=0$. There is no case 3 Killing vector for $\epsilon=0$.

 If $\epsilon=\pm 1$ we can choose $S$, $P$, \& $Q$ as described in cases 1, 2, or 3, and find a Szekeres model with one Killing vector field.  Examples of such a Killing vector field can be found in figure \ref{fig::exampleplot-cases}. If $\epsilon=0$ we can only have full symmetry or no symmetry. In particular there is no globally axial symmetric model for $\epsilon=0$ apart from full symmetry.
 
 \begin{figure}[h]
    \centering
    \subfloat[\label{KVcase1} Case 1]{
      \includegraphics[height=3cm]{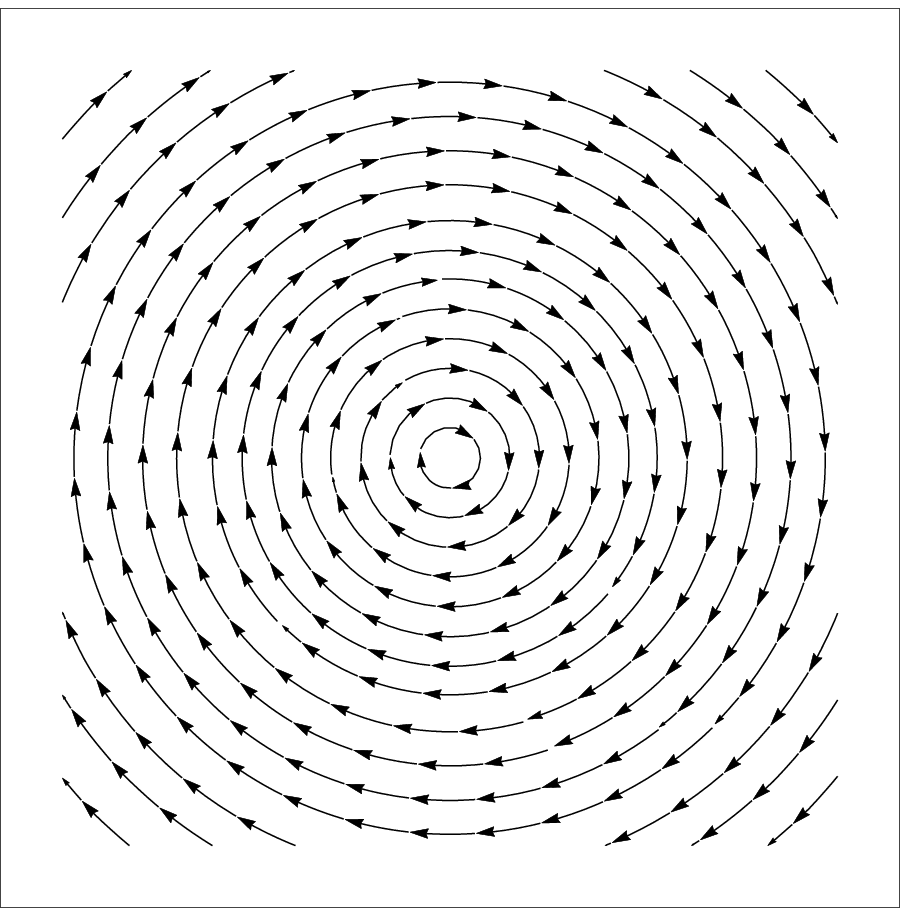}
    }
    \subfloat[\label{KVcase2a} Case 2, $Q'=0$]{
      \includegraphics[height=3cm]{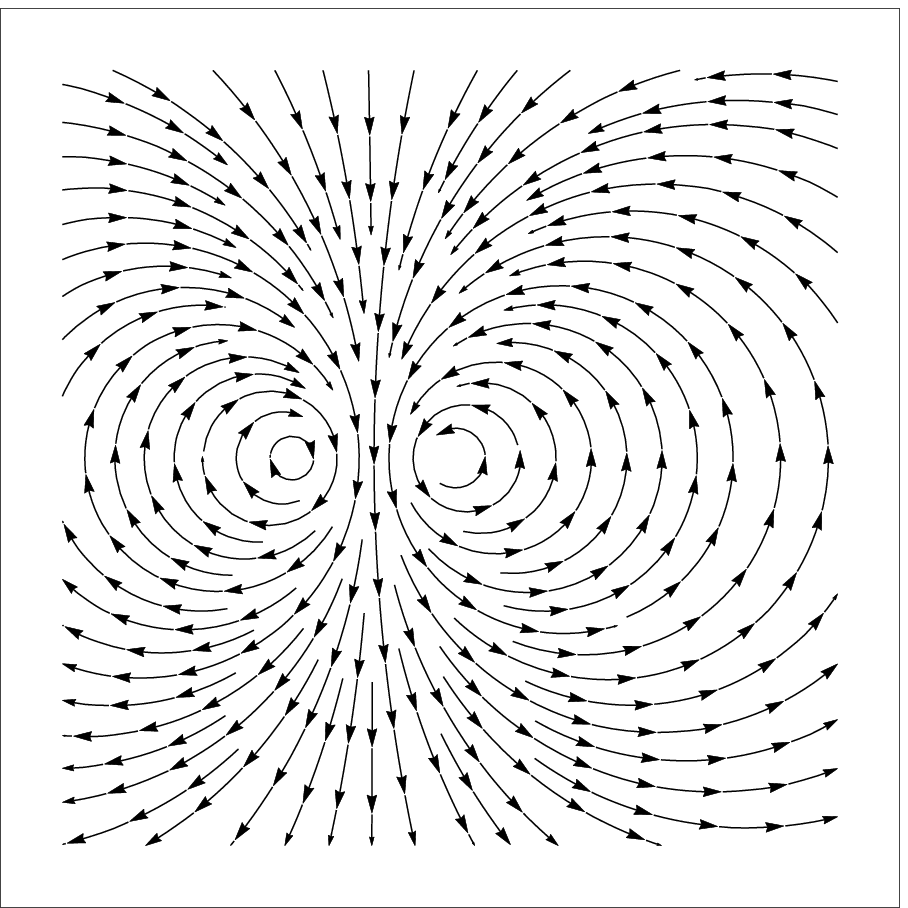}
    }
    \subfloat[\label{KVcase2b} Case 2, $P'=0$]{
      \includegraphics[height=3cm]{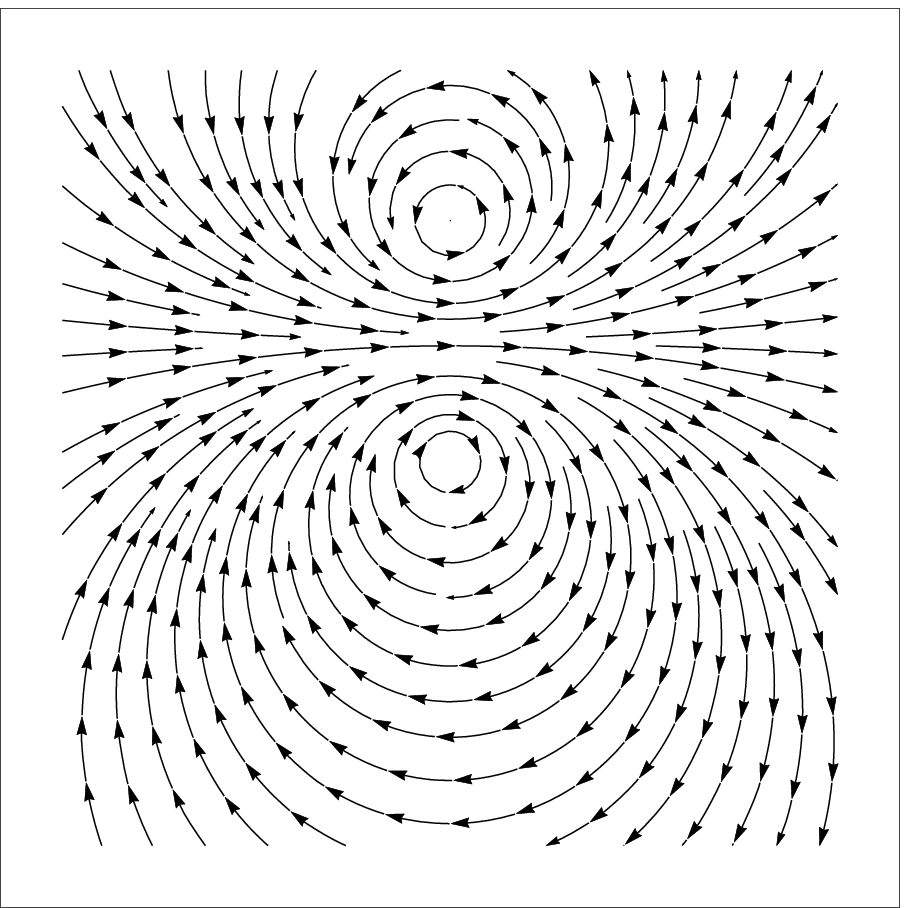}
    }
    \subfloat[\label{KVcase3} Case 3]{
      \includegraphics[height=3cm]{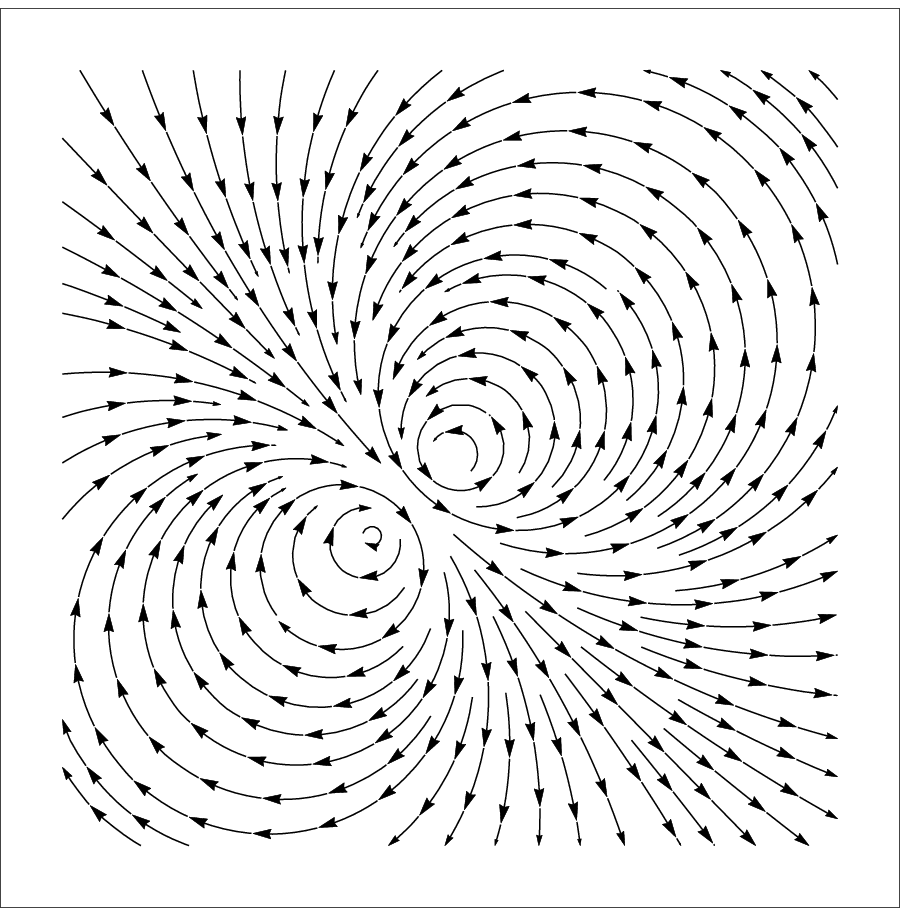}
    }
 	\caption{Examples of Killing vector fields for $\epsilon=\pm 1$.
 The plots show the stream lines in the $(p, q)$ plane. The Killing vector field is the same for every constant $(t,r)$ surface in a given symmetric model. It is clear that cases 2 and 3 are equivalent; they can be transformed to one another by a simple rotation of the $p,q$ plane}
 	\showlabel{fig::exampleplot-cases}
 \end{figure}

 Refs \cite{NolDeb07,KraBol11} showed these cases can be transformed into each other.  Conditions \er{Case3Cond} are the same as eqs (B8) \& (B11) of \cite{KraBol11}.  

 \subsection{Fixed Points}
 \showlabel{FxPt}

 We now show that the fixed points coincide with the dipole extrema.  The extrema of the dipole function $\WW$ are given by \er{ErEeS}-\er{qe} which hold generally, even if no symmetry exists.
 The fixed points of a Killing vector field are the loci $(p_f,q_f)$ where
  \begin{align}
	\xi^a = 0 ~~~~~~\Ra~~~~~~ \beta = 0 = \gamma ~.
  \end{align}
 The existence of a Killing vector field demands that conditions (\ref{KVcond1})-(\ref{KVcond3}) hold, so we consider the above list of cases.
  
    {\bf Case 1:}~~ $P'=0,Q'=0$.~~  We find
  \begin{align}
    (p_e,q_e)= (P,Q) =(p_f,q_f) ~.
  \end{align}
  
  {\bf Case 2:}~~ $P'\neq 0, Q' = 0$.~~ Expressions for $(p_f,q_f)$ result from case 3 by setting $c=0$. Then $P'=0,Q'\neq 0$ results from interchanging $P$ with $Q$, i.e.\ a rotation of the $p,q$ plane by $90^\circ$.
  
  {\bf Case 3:}~~ Both $P',Q' \neq 0$.~~ In this case
 \begin{align}
   Q= cP+ c_Q ~~~~\mbox{and}~~~~ S^2 = \frac{1}{\epsilon}(-(1+c^2)P^2+2c_2P+c_3)~.   \showlabel{Case3CondEqs}
 \end{align}
 Then we find the curves $q(p)$ that have $\beta=0$ and $\gamma=0$ as follows,
 \begin{align}
  q_\beta &= \frac{1}{c}\left(\pm\sqrt{(1+c^2) p^2-2 c_2 p+c^2 c_3+c_2^2} -p +c c_Q+c_2\right) \\
  q_\gamma &= \pm\sqrt{(1+c^2) p^2-2 c_2 p-c_3}+c p+c_Q ~.
 \end{align}
   There are two pairs of intersection points for these two curves:
 \begin{align}
    p_f & = \frac{c_2 \pm \sqrt{(1+c^2)c_3+c_2^2}\;}{c^2+1} ~~~~\mbox{and}~~~~
    q_f = c_Q + \frac{c \big(c_2 \pm \sqrt{(1+c^2) c_3+c_2^2}\; \big)}{c^2+1} ~, \\
    p_f & = \frac{c_2 \pm c\sqrt{-(1+c^2)c_3 - c_2^2}\;}{c^2+1} ~~~~\mbox{and}~~~~
    q_f = c_Q + \frac{c c_2 \mp \sqrt{-(1+c^2)c_3 - c_2^2}\;}{c^2+1} ~.
 \end{align} 
 Depending on the choice of constants, one of the two pairs is complex and the other is the pair of fixed points.
 Using the case 3 conditions \er{Case3CondEqs} we find with
 \begin{align}
  \WW|_e &= \pm \frac{P'}{\epsilon S^2}\ \sqrt{d}\, , \qquad\qquad d := c_2^2+(1+c^2)c_3 \geq 0 \showlabel{FxPtDistance}\\
   \Rightarrow~~~~ (p_e,q_e) &= \left(P-\frac{\epsilon S^2}{-(1+c^2)P+c_2 \pm \sqrt{d}},\  cP+c_Q -\frac{c\, \epsilon S^2}{-(1+c^2)P+c_2 \pm \sqrt{d}}\right)\nn\\
    &= \left( \frac{(c_2\mp \sqrt{d})}{(1+c^2)},c \frac{(c_2\mp \sqrt{d})}{(1+c^2)}+c_Q\right)=(p_f,q_f) ~.
 \end{align}
 The points of extrema coincide with the fixed points if $d\geq 0$ ($\WW|_e$ does not exist for $d\leq 0$).
 It follows that the contours of $\WW$ give the congruence of an axial Killing vector field, and we can say that $\WW$ is the symmetry of an axi-symmetric Szekeres model.

 \subsection{Discussion}
 \showlabel{Dscn}
 
 The aim of this section is firstly to determine whether a model with a single symmetry is axially symmetric or otherwise and secondly to derive the restrictions on the ranges of the functions $S$, $P$ \& $Q$.

  \begin{figure}[h]
    \centering
    \subfloat[\label{KVex1} Example for Case 1]{
      \includegraphics[height=4cm]{exampleplot_case1.eps}
    }
    \subfloat[\label{KVex2} Example for $d>0$]{
      \includegraphics[height=4cm]{exampleplot_case2b}
    }
    \subfloat[\label{KVex3} Example for $d=0$]{
      \includegraphics[height=4cm]{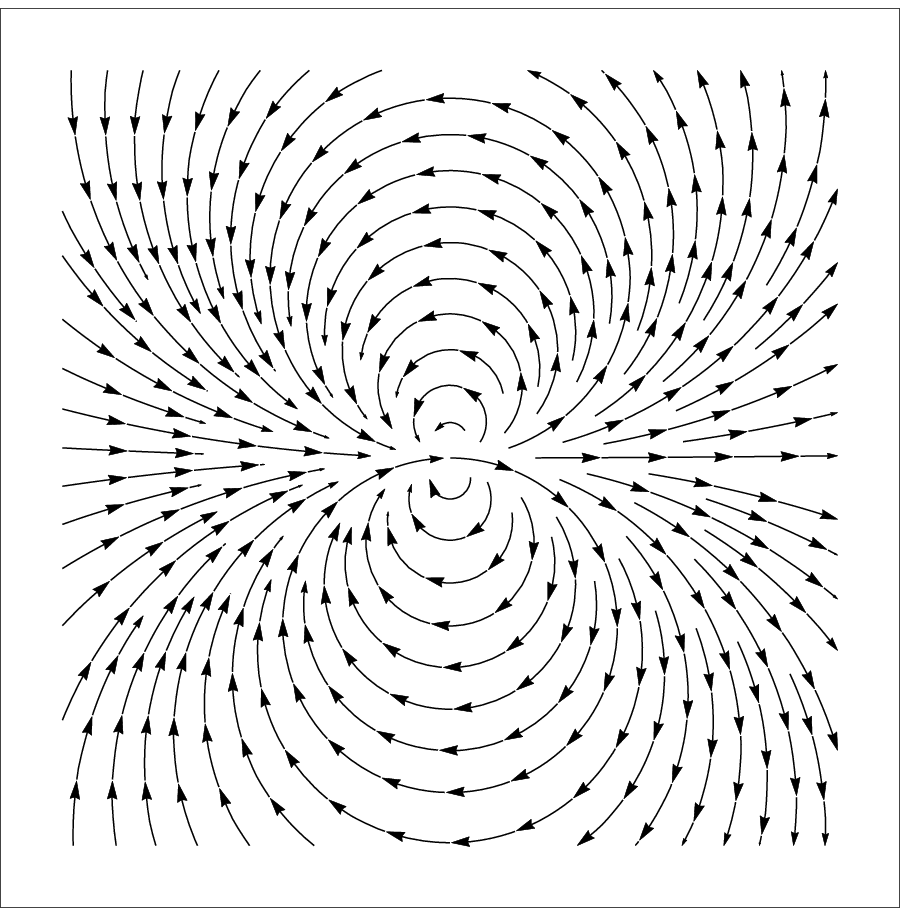}
    }
    \subfloat[\label{KVex4} Example for $d<0$]{
      \includegraphics[height=4cm]{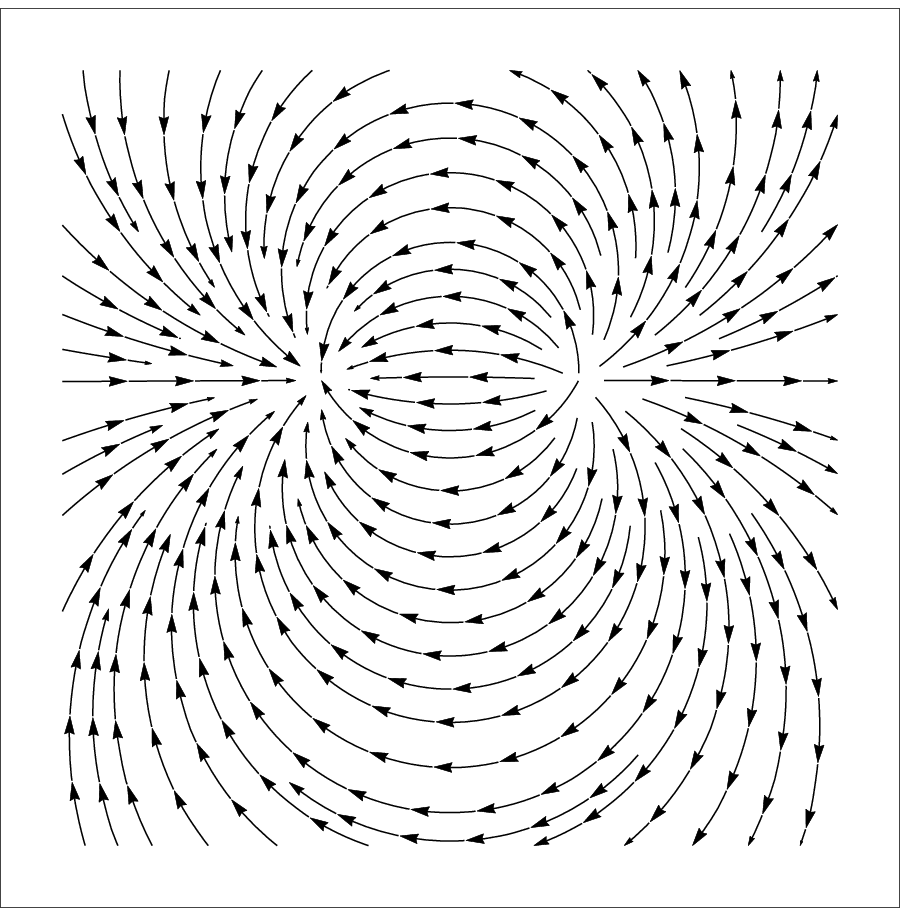}
    }
 	\caption{Examples of four possible types of Killing vector fields.  The plots show the flow in the $(p, q)$ plane. The $d$ refers to the distance between the rotational fixed points and is defined in (\ref{FxPtDistance}). The four pictures show a Killing vector field for different choices of constants $c,c_2,c_3,c_Q$.}
 	\showlabel{fig::exampleplot-types}
 \end{figure}
 
 The three cases that we found in section \ref{KlEqs} depend on the choice of the sole free function in $E$ for a model with a single symmetry. Case 1 uses $S$ as the free function, the other cases use either $P$ or $Q$ as the free funciton. The latter two cases are equivalent in the sense that a rotation of the $p,q$ plane will bring us from one to another. The really interesting difference between Killing vector fields arises from the choice of the constants $c,c_2,c_3,c_Q$. $c$ is a rotation of the coordinate system and choosing $c=0$ results in the Killing vector field aligning with the grid $(p,q)$. $c_Q$ is a simple tranlation of the coordinate system. We can set $c=c_Q=0$ without loss of generality. However, $d$ as defined in (\ref{FxPtDistance}) plays a crucial role. As shown in figure \ref{fig::exampleplot-types}, $d$ determines the type of the Killing vector field. Case 1 can be interpreted as having $d=\infty$.
 
 Now we go back and consider the different $\epsilon$ models separately. The Killing vector fields are blind to $\epsilon$, as can be seen in the explicit equations for the Killing vector components in Appendix \ref{appendix::cases}.
 But for every $\epsilon$ we have to read the Killing vector field differently. As we will see now, the four types of Killing vector fields are not all realised for each $\epsilon$, with the exception of Case 1, $d=\infty$, which represents the simplest kind of rotation. 
 
 In order to keep the equations simple we will assume case 2 ($Q'=0$). Then the fixed points of the Killing vector field become
 \begin{align}
  (p_f,q_f) &=\left( c_2\pm \sqrt{d}\,,\, c_Q\right) & \text{for } d\geq 0 ~,
  \showlabel{FxPtCase2d>0} \\
  (p_f,q_f) &=\left( c_2\,,\, c_Q\pm \sqrt{-d}\right) & \text{for } d\leq 0 ~.
  \showlabel{FxPtCase2d<0}
 \end{align}
 The line $p=c_2$ corresponds to a circle with infinite radius. If $d\geq 0$ the two fixed points are symmetric with respect to this line (see figure \ref{KVex2}). If $d\leq 0$ the two fixed points lie on this line.  Equations (\ref{FxPtCase2d>0}) and (\ref{FxPtCase2d<0}) show that the distance between the fixed points is $2\sqrt{d}$. The three different types of Killing vector fields in figure (\ref{KVex2}),(\ref{KVex3}),(\ref{KVex4}) are due to $d>0$, $d=0$ and $d<0$ respectively. The condition (\ref{Case2Cond}) can be written as
 \begin{align}
  \epsilon S^2 &= -\left(P - c_2\right)^2 + c_2^2 + c_3 ~.
  \showlabel{Case2Cond2}
 \end{align}
 Since $S > 0$, this leads to range restrictions on $P$ and $S$.
 \paragraph{$\bs{\epsilon = 1}$;}
 We can only have Killing vector fields like figure (\ref{KVex1}) and (\ref{KVex2}).The reason is simply that we would find a $S^2<0$, if $d=0$ or if $ d < 0$.
 The two types of Killing vector fields are equivalent in the sense that a conformal coordinate transformation of the $p,q$ plane can transform one into the other. They are axially symmetric models, because any Killing vector field on a sphere with two fixed points is axial symmetric. (in figure \ref{KVex1} the second fixed point is at infinity, i.e. at the north pole of the sphere.)
 The ranges of the functions $P,S$ are bounded by
 \begin{align}
  c_2-\sqrt{d} < P < c_2+\sqrt{d}, \qquad S^2 \leq d ~.
 \end{align}
 If we think of $p,q$ as coordinates on the sphere, the point $(P,Q)$ corresponds to the south pole of the sphere. It can only be between the fixed points on the line joining them. Intuitively we would have guessed that, since the fixed points must be antipodal on the sphere.
 \paragraph{$\bs{\epsilon=-1}$;}
 The quasi-hyperboloidal models can have all four types of Killing vector fields. Figures \ref{KVex1} and \ref{KVex2} are hyperbolic rotations, figure \ref{KVex4} is a hyperbolic translation, and \ref{KVex3} is a limit-rotation or horolation%
 \footnote{Since the orbits are open, this is technically a translation.}%
 .
 A hyperbolic 2-surface can be represented by the Poincar\'e disk on the $p,q$ plane (or equivalently by the upper of a two-sheeted hyperboloid). The center of the Poincar\'e disk (or the ``south pole" of the hyperboloid sheet) is $(P,Q)$, the radius of the Poincar\'e disk is proportional to $S$ (the projection height).
 If $d>0$ we find $(P,Q)$ on the line of the two fixed points but outside them, not in between. There is one fixed point on each sheet of the hyperboloid. If $d\leq 0$ the fixed points are at infinity, which is represented by the edge of the Poincar\'e disk.
 \begin{align}
   d>0&& &P< c_2-\sqrt{d} ~\text{ or }~ P > c_2+\sqrt{d}, & S^2 & > 0 ~,
   \showlabel{HypAxSym} \\
   d=0&& &P\neq c_2, & S^2 & > 0 ~,
   \showlabel{HypHor} \\
   d<0&& &P\text{ unbounded}, & S^2&\geq -d  ~.
   \showlabel{HypTransSym}
 \end{align}
  Thus we can only speak of axial symmetry if the conditions (\ref{HypAxSym}) hold.  Conditions \er{HypHor} \& \er{HypTransSym} violate \er{QHSzCondit}, and only apply in models with shell crossings somewhere.
  \paragraph{$\bs{\epsilon=0}$;}
 For purely $\epsilon=0$ models, there is either no symmetry or full planar symmetry. We cannot construct such a model with a single symmetry. 

 We have seen that the Killing vector types of figures \ref{KVex4} \& \ref{KVex3} can only occur for $\epsilon = -1$, while those of \ref{KVex1} \& \ref{KVex2} can occur for either $\epsilon = +1$ or $-1$.  It therefore seems possible that the latter two types can occur for a single $\epsilon = 0$ worldsheet with $\epsilon$ flipping across it.

 \subsection{Radial Null Geodesics}

 Nolan and Debnath \cite{NolDeb07} found that the existence of a radial null geodesic in an $\epsilon = +1$ model demands the existance of a Killing vector field.  Their work was generalised to all $\epsilon$ by Krasi\'{n}sky and Bolejko \cite{KraBol11}, who also found that only this family of symmetric Szekeres models has repeatable light rays.
 
 The equation following (31) in \cite{NolDeb07} shows that along a radial geodesic
 \begin{align}
  \lambda_{,2}\Big|_g = \lambda_{,3}\Big|_g =0 ~,
  \showlabel{lambda-geodesic}
 \end{align}
 which is true for all $\epsilon$ as they use the most general form of the geodesic equations. From our perspective this result is not surprising because
 \begin{align}
  \lambda_{,3}&\propto (ZA'-Z'A) = \frac{\beta}{h}  && ~~~~\mbox{and}~~~~ &
  \lambda_{,2}&\propto (YA'-Y'A) = \frac{\gamma}{h} ~.
  \showlabel{not-surprising}
 \end{align}
 Thus $\lambda_{,2},\lambda_{,3}$ must vanish at points $(p,q)$ where $\beta, \gamma$ vanish. Furthermore, \cite{NolDeb07} show that their $\mathcal{Q}_1=0$ and $\mathcal{Q}_2=0$ is equivalent to (\ref{lambda-geodesic}). Comparing with (\ref{BtSln1}) and (\ref{GmSln1}) shows that $\mathcal{Q}_1=\gamma$ and  $\mathcal{Q}_2=\beta$.
 
 $\epsilon=0$ is special in this regard. We can find a radial geodesic but not a single Killing vector field. The reason is, that if a radial geodesic exists then, according to \cite{NolDeb07}, we can find a coordinate transformation so that $P,Q=0$. And that, for $\epsilon=0$, leeds to $\lambda_2,\lambda_3=0$ everywhere. We find the full planar symmetry and not just a single symmetry.
 
 To sum up, a radial null geodesic exists if and only if $\beta$ and $\gamma$ are not globally vanishing, i.e.\ there exists a Killing vector.  The locus of fixed points coincides with the path of the radial geodesic.

 In fact, it is a general theorem that the locus of fixed points of a symmetric space(time) is a totally geodesic sub-space; see p224 of \cite{Helg78}, or theorem 1.3 of \cite{MAPtgs}, or p 21 of \cite{KennardMath260P}.  In our case, the geodesics in the subspace are, the symmetry axis on each time slice, the timelike worldlines, and the null geodesics along the axis.

 \section{Locus of Dipole Extrema As Geodesics}
 \showlabel{ExtrmGd}

 In the case of axial symmetry, we expect that the symetry axis should be ``straight" in the 3-d space; that is, its tangent vector should obey the 3-d geodesic equation.  Obviously it can't obey the 4-d geodesic equation, since it is well known that even for the RW metric the spacelike geodesics do not lie in constant $t$ 3-surfaces.
 We now want to ask a slightly different question than before. We do not assume that a symmetry exists but ask what models have a locus of extrema $(p_e(r),q_e(r))$ that is a spatial geodesic.

 Along the locus of extrema, $p_e$ \& $q_e$ as defined in (\ref{pe}) and (\ref{qe}) ensure that
 \begin{align}
   \left( E E'_p - E' E_p \right) |_e = 0 = \left( E E'_q - E' E_q \right) |_e   \showlabel{NoTgAc}
 \end{align}
 
 Writing the geodesic tangent vector as $V^a = [k(v), \ell(v), m(v), n(v)]$, where $v$ is a parameter, the 4-d geodesic equations in the Szekeres metric, and the spacelike condition are:
 \begin{align}
   0 & = \td{k}{v} + \frac{\ell^2}{W^2} \left( R' - \frac{R E'}{E} \right) \left( \Rt' - \frac{\Rt E'}{E} \right)
         + \frac{(m^2 + n^2) R \Rt}{E^2} ~,   \showlabel{Gdt} \\
   0 & = \td{\ell}{v} + 2 \ell k \frac{(\Rt' - \Rt E'/E)}{(R' - R E'/E)}
         + \ell^2 \left( \frac{(R'' - R E''/E)}{(R' - R E'/E)} - \frac{E'}{E} - \frac{W'}{W} \right) \nn \\
      &~~~~ + 2 \ell R \left( \frac{m (E' E_p - E E'_p) + n (E' E_q - E E'_q)}{E^2 (R' - R E'/E)} \right)
         - \frac{(m^2 + n^2) R W^2}{E^2 (R' - R E'/E)} ~,
         \showlabel{Gdr} \\
   0 & = \td{m}{v} + \frac{2 k m \Rt}{R} + \frac{\ell^2}{R W^2} \left( R' - \frac{R E'}{E} \right)
         \left( E E'_p - E' E_p \right) \nn \\
      &~~~~ + \frac{2 \ell m}{R} \left( R' - \frac{R E'}{E} \right) - \frac{(m^2 - n^2) E_p + 2 m n E_q}{E} ~,
         \showlabel{Gdp} \\
   0 & = \td{n}{v} + \frac{2 k n \Rt}{R} + \frac{\ell^2}{R W^2} \left( R' - \frac{R E'}{E} \right)
         \left( E E'_q - E' E_q \right) \nn \\
      &~~~~ + \frac{2 \ell n}{R} \left( R' - \frac{R E'}{E} \right) - \frac{(n^2 - m^2) E_q + 2 m n E_p}{E} ~,
         \showlabel{Gdq} \\
   + 1 & = - k^2 + \frac{\ell^2}{W^2} \left( R' - \frac{R E'}{E} \right)^2 + \frac{(m^2 + n^2) R^2}{E^2} ~.  \showlabel{SpL}
 \end{align}
 The 3-d geodesics equations are obtained by setting
 \begin{align}
    k = 0 = \Rt = \Rt' ~.   \showlabel{Gd3dCond}
 \end{align}

 We will now show that, provided the non-spherical arbitrary functions $S$, $P$ \& $Q$ have no dependence on the spherical arbitrary functions $f$, $M$ \& $t_B$, then the locus of dipole extrema is only geodesic if it is along constant $(p, q)$.

 Since $p_e$ \& $q_e$ are functions of $r$ only, we write the tangent vector to this locus as
 \begin{align}
   V^b & =  \left[ 0, 1, p_e'(r), q_e'(r) \right] \ell(r) ~.   \showlabel{PlLcTnVc}
 \end{align}
 Along the pole locus, \er{NoTgAc} \& \er{PlLcTnVc} reduce the 3-d geodesic equations, \er{Gdr}-\er{SpL} with \er{Gd3dCond}, to
 \begin{align}
   0 & = \ell^2 \left\{ \frac{\ell'}{\ell} + \frac{(R''/R - E''/E)}{(R'/R - E'/E)} - \frac{E'}{E} - \frac{W'}{W}
      - \frac{(p_e'^2 + q_e'^2) W^2}{E^2 (R'/R - E'/E)} \right\} ~,
      \showlabel{Gd3rPl} \\
   0 & = \ell^2 \left\{ p_e'' + p_e' \left[ \frac{\ell'}{\ell} + 2 \left( \frac{R'}{R} - \frac{E'}{E} \right) \right]
      - \frac{(p_e'^2 - q_e'^2) E_p + 2 p_e' q_e' E_q}{E} \right\} ~,
      \showlabel{Gd3pPl} \\
   0 & = \ell^2 \left\{ q_e'' + q_e' \left[ \frac{\ell'}{\ell} + 2 \left( \frac{R'}{R} - \frac{E'}{E} \right) \right]
      - \frac{(q_e'^2 - p_e'^2) E_q + 2 p_e' q_e' E_p}{E} \right\} ~,
      \showlabel{Gd3qPl} \\
   + 1 & = \ell^2 R^2 \left\{ \frac{(R'/R - E'/E)^2}{W^2} + \frac{(p_e'^2 + q_e'^2)}{E^2} \right\} ~.
      \showlabel{SpL3Pl}
 \end{align}
 Now $E$ and its derivatives depend on $r$ through $S$, $P$ \& $Q$, as do $p_e$ \& $q_e$, while $R'/R$ depends on $r$ through $f$, $M$ \& $t_B$.  But, both \er{Gd3rPl} and \er{SpL3Pl} show ~$\ell'/\ell \neq - 2 (R'/R - E'/E)$~, even if $p_e = 0 = q_e$.  Consequently, equations \er{Gd3pPl} \& \er{Gd3qPl} show that for $(p_e, q_e)$ to be geodesic, either $E'/E$ depends on $R'/R$, which we have excluded, or
 \begin{align}
   p_e'|_\text{geodesic} = 0 = q_e'|_\text{geodesic} ~.
 \end{align}
 Therefore the only way that the locus of dipole extrema can be made geodesic without putting restrictions on the spherical arbitrary functions, is if $p_e$ and $q_e$ are constant.  Setting \er{pe} \& \er{qe} constant results in the same set of cases and restrictions on $S$, $P$ \& $Q$ as in section \ref{KlEqs}.

 \section{Equivalences Between Szekeres Metrics}
 \showlabel{SzEquiv}

 The set of axi-symmetric Szekeres models found in section \ref{KlEqs} are known to be equivalent \cite{NolDeb07,KraBol11}.  Do there exist equivalences between general Szekeres models?

 The non-sphericity or non-pseudo-sphericity of a Szekeres model could be characterised by the variation of its dipole strength $\WW|_e$, and the path of its dipole extrema, $p_e$ \& $q_e$, all of which are functions of $r$ only.  We conjecture that two Szekeres models, with distinct sets of non-spherical functions $(S, P, Q)$, are equivalent if one can be ``rotated" (in the spherical or hyperboloidal sense) into the other.  Any coordinate transformation of a given metric is necessarily the same physical spacetime.  Therefore we are looking for the most general coordinate transformation that preserves the Szekeres form.  We already know there is a rescaling freedom in the $r$ coordinate.  Since we must retain the forms of \er{ds2Unit}, \er{Edef} \& $E'/E$, we are looking for transformations of the $(p, q)$ coordinates only which are $r$-independent.  Indirectly, this will result in a transformation of $S$, $P$, $Q$.

 The symmetries of constant curvature surfaces have been well studied.  The transformations that preserve the 2-d metric form%
 \footnote{The general transformation is often written in the form of a Mobius tranformation,
 \begin{align}
   (\tilde{p} + i \tilde{q}) = \frac{k (p + i q) + l}{m (p + i q) + n} ~,   \showlabel{Mobius}
 \end{align}
 where $k$, $l$, $m$ \& $n$ are complex constants obeying $k n - l m = 1$.
 }
 \er{ds2Unit} and \er{Edef} are composed of equatorial rotations (Haantjes transformations)
 \begin{align}
   \begin{split}
      T & = 1 + 2 D_1 p + 2 D_2 q + (D_1^2 + D_2^2)(p^2 + q^2) ~, \\
      \tilde{p} & = \frac{p + D_1 (p^2 + q^2)}{T} ~,~~~~~~
         \tilde{q} = \frac{q + D_2 (p^2 + q^2)}{T} ~,
   \end{split}   \showlabel{HaantjesER} 
 \intertext{polar rotations}
   \tilde{p} & = \frac{F_1 p - F_2 q}{\sqrt{F_1^2 + F_2^2}\;} ~,~~~~~~
      \tilde{q} = \frac{F_2 p + F_1 q}{\sqrt{F_1^2 + F_2^2}\;} ~,   \showlabel{HaantjesPR}
 \intertext{inversions}
   \tilde{p} & = - p ~,~~~~~~ \tilde{q} = q ~,   \showlabel{HaantjesI}
 \intertext{magnifications}
   \tilde{p} & = \mu p ~,~~~~~~ \tilde{q} = \mu q ~,   \showlabel{HaantjesM}
 \intertext{and displacements}
   \tilde{p} & = p_0 + p ~,~~~~~~ \tilde{q} = q_0 + q ~.   \showlabel{HaantjesD}
 \end{align}
 in any combination.

 In order for any of these transformations to preserve the form of the full Szekeres metric \er{ds2Sz}, we require that each of $(\d p^2 + \d q^2)/E^2$ and $E'/E$ are invariant under
 \begin{align}
   p ~\to~ \tilde{p} ~,~~ q ~\to~ \tilde{q} ~,~~ S ~\to~ \tilde{S} ~,~~ P ~\to~ \tilde{P} ~,~~ Q ~\to~ \tilde{Q} ~.
   \showlabel{Map}
 \end{align}
 This will result in an associated transformation of $S$, $P$, \& $Q$.

 As an example, we consider an equatorial rotation, and apply \er{Map} plus \er{HaantjesER} to $(\d p^2 + \d q^2)/E^2$ with general $E$.  Simplifying, we get
 \begin{align}
   & \frac{(\d \tilde{p}^2 + \d \tilde{q}^2)}{\tilde{E}^2} ~~~~\to~~~~ \frac{4 \tilde{S}^2 (\d p^2 + \d q^2)}{F}
      = \frac{(\d p^2 + \d q^2)}{E^2} ~, \\
   \mbox{where}~~~~ F & = \Big[ (p^2 + q^2) - 2 \big[ p \tilde{P} + q \tilde{Q}
         + (p^2 + q^2)(D_1 \tilde{P} + D_2 \tilde{Q}) \big] \nn \\
      &~~~~ + \big[ 1 + 2 D_1 p + 2 D_2 q + (p^2 + q^2)(D_1^2 + D_2^2) \big]
         (\tilde{P}^2 + \tilde{Q}^2 + \epsilon \tilde{S}^2) \Big]^2 ~.
 \end{align}
 Since we must have ~$E^2 = F/(4 \tilde{S}^2)$~, we compare coefficients of powers of $p$ \& $q$ to obtain
 \begin{align}
   \begin{split}
      \frac{1 - 2 [D_1 \tilde{P} + D_2 \tilde{Q}] + (D_1^2 + D_2^2)
         [\tilde{P}^2 + \tilde{Q}^2 + \epsilon \tilde{S}^2]}{2 \tilde{S}} & = \frac{1}{2 S} \\
      \frac{- \tilde{P} + D_1 [\tilde{P}^2 + \tilde{Q}^2 + \epsilon \tilde{S}^2]}{\tilde{S}} & = \frac{-P}{S} \\
      \frac{- \tilde{Q} + D_2 [\tilde{P}^2 + \tilde{Q}^2 + \epsilon \tilde{S}^2]}{\tilde{S}} & = \frac{-Q}{S} \\
      \frac{[\tilde{P}^2 + \tilde{Q}^2 + \epsilon \tilde{S}^2]}{2 \tilde{S}} & =
         \frac{S}{2} \left( \frac{P^2}{S^2} + \frac{Q^2}{S^2} + \epsilon \right) ~,
   \end{split}
 \end{align}
 which gives us the transformation of $S$, $P$ \& $Q$ under an equatorial rotation:
 \begin{align}
   \begin{split}
      U & = 1 + 2 D_1 P + 2 D_2 Q + (D_1^2 + D_2^2) (P^2 + Q^2 + \epsilon S^2) ~, \\
      \tilde{S}_{ER} & = \frac{S}{U} ~, \\
      \tilde{P}_{ER} & = \frac{P + D_1 (P^2 + Q^2 + \epsilon S^2)}{U} ~, \\
      \tilde{Q}_{ER} & = \frac{Q + D_2 (P^2 + Q^2 + \epsilon S^2)}{U} ~.
   \end{split}   \showlabel{SPQtransfER}
 \end{align}

 It may be confirmed that the form and value $E'/E$ is similarly preserved by the above.

 Following a similar process for the other transformations, we find
 \begin{align}
   \tilde{S}_{PR} & = S ~,~~~~~~
      \tilde{P}_{PR} = \frac{(F_1 P - F_2 Q)}{\sqrt{F_2^2 + F_2^2}\;} ~,~~~~~~
      \tilde{Q}_{PR} = \frac{(F_1 Q + F_2 P)}{\sqrt{F_2^2 + F_2^2}\;} ~,
      \showlabel{SPQtransfPR}
 \end{align}
 \begin{align}
   \tilde{S}_{I} & = S ~,~~~~~~ \tilde{P}_{I} = - P ~,~~~~~~ \tilde{Q}_{I} = Q ~,   \showlabel{SPQtransfI}
 \end{align}
 \begin{align}
   \tilde{S}_{M} & = \mu S ~,~~~~~~ \tilde{P}_{M} = \mu P ~,~~~~~~ \tilde{Q}_{M} = \mu Q ~,   \showlabel{SPQtransfM}
 \end{align}
 \begin{align}
   \tilde{S}_{D} & = S ~,~~~~~~ \tilde{P}_{D} = p_0 + P ~,~~~~~~ \tilde{Q}_{D} = q_0 + Q ~.   \showlabel{SPQtransfD}
 \end{align}

 One may also write down some discrete equivalences, such as
 \begin{align}
   (\tilde{p}, \tilde{q}, \tilde{P}, \tilde{Q}) & = (q, p, Q, P) ~, \\
   (\tilde{p}, \tilde{q}, \tilde{P}, \tilde{Q}) & = (p, -q, P, -Q) ~,
 \end{align}
 but these are combinations of inversions \er{HaantjesI}+\er{SPQtransfI} and specific polar rotations \er{HaantjesPR}+\er{SPQtransfPR}.  Since the sign of $S$ does not affect the metric (provided $S$ is never zero), there is also the trivial equivalence
 \begin{align}
   S \to - S ~.
 \end{align}

 Therefore equivalent Szekeres metrics, whether or not they have symmetries, are related by \er{Map}, with \er{HaantjesER} \& \er{SPQtransfER} or \er{HaantjesPR} \& \er{SPQtransfPR} or \er{HaantjesI} \& \er{SPQtransfI} or \er{HaantjesM} \& \er{SPQtransfM} or \er{HaantjesD} \& \er{SPQtransfD}, and these may be combined in the obvious way.  A similar list of $(p, q)$ transformations with their effects on $S$, $P$, \& $Q$, is given in \cite{KraBol11} equations (B12)-(B20), though magnifications appear to be missing.  The transformations \er{Mobius} \& \er{HaantjesER} with \er{HaantjesD} were used in \cite{NolDeb07,BoKrHeCe10,KraBol11} to show that certain axially symmetric models are equivalent.  However, \er{Mobius} does not provide the relationships between the two sets of $S$, $P$, \& $Q$ functions in equivalent models.

 \section{Conclusion}

 We have considered the conditions under which symmetries do and don't exist in Szekeres metrics, by attempting to solve the Killing equations and finding all circumstances that make this possible.

 The theorem of Bonnor, Suleiman and Tomimura, that the Szekeres metric has no Killing vectors if the free functions $f$, $M$, $t_B$, $S$, $P$, $Q$ are linearly independent, is easily extended from $\epsilon = +1$ to all $\epsilon$ values.  Consequently, it is immediately obvious that restrictions on the free functions are needed for Killing vectors to exist.  

 By insisting a solution to the Killing equations exists, the general solution forms \er{BtSln1}-\er{GmSln2} were found, which are valid for all $\epsilon$, if they can be made independent of the $r$ coordinate.  However, the resulting restrictions \er{KVcond1}-\er{KVcond3} on the functions $S$, $P$ and $Q$, divide into 3 cases, which restrict their possible ranges.  In each non-trivial case, $S$, $P$ and $Q$ depend on a single free function, as shown by \er{PQSint}, and in the simplest case this is obvious.  The exact nature of the restrictions depends on the value of $\epsilon$.

 The $\epsilon = +1$ Szekeres models are foliated by $(p,q)$ 2-surfaces that are spheres.  All symmetries are rotations about an axis that is common to every constant $t,r$ shell.

 Szekeres models with $\epsilon = -1$ \cite{HelKra08} are foliated by $(p,q)$ 2-surfaces that can be thought of as hyperbolids of revolution that have two possible sheets.  The important condition \er{QHSzCondit} separates models which have no ``dipole" from those which do.  Similarly, shell crossings can only be avoided in the models with a ``dipole" and only in one of the sheets.  In such models, the Killing vectors may describe rotations or translations, but interestingly the rotations only exist in models that contain a ``dipole", and the translations only exist in models that don't.  The rotations have fixed points that coincide with the poles of the ``dipole" --- the extremes of $E'/E$ on each $(p,q)$ 2-surface, while the fixed points of the translations do not.

 In the $\epsilon = 0$ Szekeres models, the foliating 2-surfaces are planes.  It turns out that the functional restrictions force full planar symmetry, so these models cannot have just one symmetry.  Interestingly, they cannot be globally free of shell crossings unless there's full planar symmetry.

 While the set of axially symmetric solutions was already known \cite{NolDeb07,KraBol11}, as the answer to two other questions, the full set of single-symmetry Szekeres models had not been looked for, and it wasn't known if the set was complete, till now.  The same references also showed the 3 cases for each $\epsilon \neq 0$ are equivalent since they can be transformed into each other.

 There is a theorem that in any spacetime with a symmetry, the locus of fixed points is a totally geodesic submanifold.  In particular, with axial symmetry, the locus of fixed points, that is the axis, must be ``straight", i.e. geodesic in the constant $t$ 3-spaces.  For Szekeres models we found that the dipole extrema also lie on this locus.  In section \ref{ExtrmGd}, we asked a slightly different question, and showed that if the locus of dipole extrema is required to be geodesic, then we get the same family of symmetric Szekeres models.

 In section \ref{SzEquiv} we have presented the full set of equivalences between general Szekeres models.  If two Szekeres metrics have the same $f$, $M$ and $a$, and their $S$, $P$ \& $Q$ are related by one of these equivalences, or a combination of them, then they are physically equivalent, and can be transformed into each other.  The list in \cite{KraBol11} was not quite complete.

 \appendix

 \section{BST Revisited}
 \showlabel{BSTKV}

 We here review the salient part of \cite{BoSuTo77} and comment on generalising to $\epsilon\neq 1$.  Assuming a non-vanishing Killing vector field of the form \er{KVform}, we find the Killing vector equations \er{KlngVcDf} for metric \er{ds2SzGen} to be
 \begin{align}
   & 11 & \alpha_{,1} + \frac{1}{2} \big( \lambda_{,1} \, \alpha + \lambda_{,2} \, \beta + \lambda_{,3} \, \gamma
      + \lambda_{,0} \, \delta \big) & = 0   \showlabel{Ke11} \\
   & 12 & e^\lambda \, \alpha_{,2} + e^\omega \, \beta_{,1} & = 0   \showlabel{Ke12} \\
   & 13 & e^\lambda \, \alpha_{,3} + e^\omega \, \gamma_{,1} & = 0   \showlabel{Ke13} \\
   & 10 & \delta_{,1} - e^\lambda \, \alpha_{,0} & = 0   \showlabel{Ke01} \\
   & 22 & \beta_{,2} + \frac{1}{2} \big( \omega_{,1} \, \alpha + \omega_{,2} \, \beta + \omega_{,3} \, \gamma
      + \omega_{,0} \, \delta \big) & = 0   \showlabel{Ke22} \\
   & 23 & \beta_{,3} + \gamma_{,2} & = 0   \showlabel{Ke23} \\
   & 20 & \delta_{,2} - e^\omega \, \beta_{,0} & = 0   \showlabel{Ke02} \\
   & 33 & \gamma_{,3} + \frac{1}{2} \big( \omega_{,1} \, \alpha + \omega_{,2} \, \beta + \omega_{,3} \, \gamma
      + \omega_{,0} \, \delta \big) & = 0   \showlabel{Ke33} \\
   & 30 & \delta_{,3} - e^\omega \, \gamma_{,0} & = 0   \showlabel{Ke03} \\
   & 00 & \delta_{,0} & = 0 ~.   \showlabel{Ke00}
 \end{align}
 The comoving matter flow lines are geodesic, so their Lie derivative with respect to $\xi^a$ must vanish
 \begin{align}
   \xi^i{}_{;k} \, u^k - \xi^k \, u^i{}_{;k} = 0 = \xi^i{}_{,0} ~,
 \end{align}
 showing $\xi^a$ is not dependent on the time coordinate.  From \er{Ke01}, \er{Ke02}, \er{Ke03} and \er{Ke00} we find that
 \begin{align}
   0 = \delta_{,1} = \delta_{,2} = \delta_{,3} ~~\to~~ \delta ~\mbox{is constant} ~.   \showlabel{DeltaConst}
 \end{align}
 For each of \er{Ke12} and \er{Ke13}, multiplying by $e^{-\omega}$, taking the derivative with respect to $t$, and applying \er{Ke12} gives
 \begin{align}
   \alpha_{,2} \, (\omega_{,0} - \lambda_{,0}) e^{\lambda - \omega} = 0 = 
      \alpha_{,3} \, (\omega_{,0} - \lambda_{,0}) e^{\lambda - \omega} ~,
    \showlabel{alpha2alpha3}
 \end{align}
 but $e^{\lambda-\omega}=0$ corresponds to a singularity in the metric and the mass density \er{MassDensity}. Furthermore,
 \begin{align}
 \omega_{,0} - \lambda_{,0} = \frac{R}{R'-R\WW}\ \omega_{,01} = \frac{2 R}{R'-R\WW} \left(\frac{\dot{R}}{R}\right)'
 \end{align}
 and $\omega_{,01}=0$ corresponds to a FLRW model, because it results in $R(t,r)= a(t)\cdot \Phi(r)$, which \er{fnotM32} disallows.
 Thus, by \er{alpha2alpha3}, \er{Ke12} and \er{Ke13} we have
 \begin{align}
   \alpha_{,2} &= 0 = \alpha_{,3}\ ,&& ~~~~\mbox{and}~~~~ &
   \beta_{,1} &= 0 = \gamma_{,1} ~.   \showlabel{BtGmIndepr}
 \end{align}
 We next take the $r$ derivative of \er{Ke22}, apply \er{DeltaConst} \& \er{BtGmIndepr},
 \begin{align}
   0 & = \frac{1}{2}  \big( \omega_{,1} \, \alpha_{,1} + \omega_{,11} \, \alpha + \omega_{,21} \, \beta
      + \omega_{,31} \, \gamma + \omega_{,01} \, \delta \big) ~,
 \end{align}
 and substitute for $\alpha_{,1}$ from \er{Ke11}
 \begin{align}
   \alpha_{,1} = - \frac{1}{2} \big( \lambda_{,1} \, \alpha + \lambda_{,2} \, \beta + \lambda_{,3} \, \gamma
      + \lambda_{,0} \, \delta \big) ~,
 \end{align}
 to get
 \begin{align}
   0 & = \alpha \left( \omega_{,11} - \frac{1}{2} \lambda_{,1} \, \omega_{,1} \right)
         + \beta \left( \omega_{,12} - \frac{1}{2} \lambda_{,2} \, \omega_{,1} \right) \nn \\
      &~~~~ + \gamma \left( \omega_{,13} - \frac{1}{2} \lambda_{,3} \, \omega_{,1} \right)
         + \delta \left( \omega_{,10} - \frac{1}{2} \lambda_{,0} \, \omega_{,1} \right) ~.
 \end{align}
 Noting that
 \begin{align}
   \left( \omega_{,11} - \frac{1}{2} \lambda_{,1} \, \omega_{,1} \right) & = 
      - 2 \left( \frac{R'}{R} - \frac{E'}{E} \right) \left( \frac{R'}{R} - \frac{W'}{W} \right) \\
   \left( \omega_{,12} - \frac{1}{2} \lambda_{,2} \, \omega_{,1} \right) & = 0 =
      \left( \omega_{,13} - \frac{1}{2} \lambda_{,3} \, \omega_{,1} \right) \\
   \left( \omega_{,10} - \frac{1}{2} \lambda_{,0} \, \omega_{,1} \right) & = 
      - \frac{2 \Rt}{R} \left( \frac{R'}{R} - \frac{E'}{E} \right) ~,
 \end{align}
 this reduces to
 \begin{align}
   0 &= - 2 \left( \frac{R'}{R} - \frac{E'}{E} \right) \frac{1}{R} \left\{ \alpha \left( R' - R\, \frac{W'}{W} \right)
         + \delta \Rt \right\}  ~.
 \end{align}
 Again we don't allow $e^{\lambda-\omega} =( \frac{R'}{R} - \frac{E'}{E} )^2(\frac{E}{W})^2$ to vanish, in order to avoid singularities, and therefore
 \begin{align}
	0 &= 
	\alpha\left(R'- R\,\frac{W'}{W}\right) +\delta\dot{R} ~.
	\showlabel{AlDtEq}
 \end{align}
 After a derivation with respect to $t$ and multiplying with $\dot{R}$, we can use the Friedman equation \er{RtSqEq} and its $t$ and $r$ derivatives to eliminate $\Rt^2,\ddot{R},\dot{R}'$. In addition we use Equation \er{AlDtEq} to eliminate $\delta$. Then,
 \begin{align}
	0 
	&= \alpha\left(-\epsilon\frac{W'}{W}-\frac{M'}{R}
	+3\frac{W'}{W}\frac{M}{R}\right) ~.
 \end{align}
 Since $R$ is time dependent and $M,W$ are not, the bracket term can only vanish if the following two equations are satisfied simultaneously:
 \begin{align}
	\epsilon \frac{W'}{W} &=0, 
	& M'-3\frac{MW'}{W} &=0 ~.
	\showlabel{MWLinDep}
 \end{align}
 At this point BST assumed $\epsilon = +1$ in order to show that $\alpha = 0$.  However the result follows for all $\epsilon$, 
if we alter the assumption regarding $M,f$ slightly. 
 For $\epsilon \neq 0$ we must have $W'=0$ and thus also $M'=0$. Clearly this is not an interesting model%
 \footnote{If $M' = 0$ we have a vacuum model, which forces spherical symmetry to avoid shell crossings \cite{HelKra02}.}%
 , and by assumption \er{fnotM32}, we exclude this choice of special LT-functions.

 For $\epsilon =0$ we only have
 \begin{align} 
	\frac{M'}{M} &= 3\frac{W'}{W} 
	&&\Rightarrow\quad & M &= \text{const}\cdot f^{3/2} ~,
 \end{align}
which we again exclude by \er{fnotM32}.%
 \footnote{Note that $\frac{f}{M^{2/3}}=$const. allows for a constant $t$-component $\delta$ and an $r$-component $\alpha(r)$ of the Killing vector.}
 Therefore we must have $\alpha=0$, and from equation (\ref{AlDtEq}) it follows that $\delta=0$.
 The Killing vector now has the form \er{BtGmpq}, which is used in section \ref{KlEqs}.

 BST go on to show that either there must be a linear relation between the three functions $1/(2 S)$, $-P/(2 S)$, $-Q/(2S)$, or the Killing vector must vanish everywhere.  This means a necessary condition for a non-vanishing Killing vector field is
 \begin{align}
   0 = \frac{1}{(2 S)} \big( \sigma_1 - \sigma_2 P - \sigma_3 Q \big)
   ~~~~~~\Ra~~~~~~ P = c Q + c_Q ~,
 \end{align}
 where $\sigma_i$, $c$, $c_Q$ are constants, not all zero. This condition emerged in \er{PQSint} and \er{Case3Cond}.

 {\it Remark:}~ Strictly speaking, BST only proved their theorem for $\epsilon=1$, but they used general expressions for most of their proof, so it is not difficult to generalise it.  They used the restriction $\epsilon =1$ on only two occasions. 
 First, it is used to show that in their notation $a(r)\neq 0$, which is needed in deriving their eq (4.22). But this is true for all $\epsilon$ models, because there exists a coordinate transformation in the $p,q$ surface so that $a\neq 0$ even if $a=0$ initiallly. The proof can be found in Plebanski \& Krasinski \cite{PleKra06}. Because of this, we are able to write $a=\frac{1}{2S}$ and our notation is well defined.
 Second, as mentioned above, they used $\epsilon=1$ to show $\delta=0=\alpha$ using \er{MWLinDep}. A slight correction in the assumption regarding $M,f$ will result in the theorem being true for all $\epsilon$.
 The theorem statement for all $\epsilon$ then is:
 
 \paragraph{Generalised BST Theorem}
 Consider a Szekeres space-time (of class I) satisfying the following conditions
 \begin{align}
  e^{\lambda-\omega} &\neq 0 ~, \\
  R\neq 0,~~ E &\neq 0,~~ W=\epsilon+f \neq 0 ~, \\
  \omega_{,01} &\neq 0 ~, \\
  \text{for } \epsilon=\pm 1 &\text{ : } M'\neq 0,~~ W' = f' \neq 0 ~, \\
  \text{for } \epsilon= 0 &\text{ : } \frac{f }{ M^{2/3}}\neq \text{constant} ~. \\
  \frac{1}{2S},~ -\frac{P}{2S},~ -\frac{Q}{2S} &\text{ are linearly independent} ~,
 \end{align}
 then $\xi^\mu=0$ except possibly on isolated submanifolds of the space-time.
 In the first 3 lines above, zeros may occur at restricted loci, but not in general.

 \section{Details of the Cases}
 \showlabel{appendix::cases}

 We here collect the results for the cases listed in section \ref{SzSingSymm}.

 \paragraph{Case 1:}~~ $P'=0$, $Q'=0$.
 \begin{align}
   \mbox{Function conditions} &&& P, Q ~\mbox{constant} \\
   \mbox{Killing field} &&& \beta = 2 c_s (q - Q) \\
   &&& \gamma = - 2 c_s (p - P) \\
   \mbox{Fixed points} &&& p_f = P ~,~~~~ q_f = Q
 \end{align}
 There is only one fixed point in the $(p, q)$ plane, the other is the circle at infinity. The Killing vector field consists of concentric circular orbits, as in figure \ref{KVex1}.

 \paragraph{Case 2a:}~~ $P'=0$, $Q'\neq0$.
 \begin{align}
   \mbox{Function conditions} &&& \epsilon S^2 = - Q^2 + 2 c_2 Q + c_3 \\
   \mbox{Killing field} &&& \beta = c_q ((p - P)^2 - q^2 + 2 c_2 q + c_3) \\
   &&& \gamma = 2 c_q (p - P) (q - c_2) \\
   \mbox{Fixed points} &&& p_f = P ~,~~~~~~ q_f = c_2 \pm \sqrt{c_2^2 + c_3}\; ~,~~~~~~ c_3 \geq - c_2^2 \\
   \mbox{or} &&& p_f = P \pm \sqrt{-c_2^2 - c_3}\; ~,~~~~~~ q_f = c_2 ~,~~~~~~ c_3 \leq - c_2^2 ~,~~ \epsilon = -1
 \end{align}
 The first fixed point pair lies on the vertical line $p = P$, and the second pair lies on the circle ~$Y^2 + Z^2 = Q^2 - 2 c_2 Q - c_3 = S^2$~.  

 \paragraph{Case 2b:}~~ $P'\neq0$, $Q'=0$.
 \begin{align}
   \mbox{Function conditions} &&& \epsilon S^2 = - P^2 + 2 c_2 P + c_3 \\
   \mbox{Killing field} &&& \beta = 2 c_p (q - Q) (c_2 - p) \\
   &&& \gamma = c_p (p^2 - (q - Q)^2 - 2 c_2 p - c_3) \\
   \mbox{Fixed points} &&& p_f = c_2 \pm \sqrt{c_2^2 + c_3}\; ~,~~~~~~ q_f = Q ~,~~~~~~ c_3 \geq - c_2^2 \\
   \mbox{or} &&& p_f = c_2 ~,~~~~~~ q_f = Q \pm \sqrt{-c_2^2 - c_3}\; ~,~~~~~~ c_3 \leq - c_2^2 ~,~~ \epsilon = -1
 \end{align}
 The Killing vector field is as above, but rotated by $90^\circ$ in the $(p, q)$ plot, as in fig \ref{KVex4}.

 \paragraph{Case 3:}~~ $P'\neq0$, $Q'\neq0$.
 \begin{align}
   \hspace*{-4mm}\mbox{Function conditions} &&& Q = c P + c_Q ~,~~~~~~ \epsilon S^2 = - (1 + c^2) P^2 + 2 c_2 P + c_3 \\
   \mbox{Killing field}
      &&& \beta = c_p \Big[ (p^2 - q^2) c - 2 p q + 2 q (c c_Q + c_2) + 2 p c_Q + (cc_3 - c c_Q^2 - 2 c_2 c_Q) \Big] \\
   &&& \gamma = - c_p \Big[ - p^2 + q^2 - 2 c p q + 2 p (c_Q + c_2) + 2 q c_Q + c_Q^2 + c_3 \Big] \\
   \mbox{Fixed points} &&& \beta=0 ~~~~\Ra~~~~ q = \frac{c c_Q + c_2 - p
      \pm\sqrt{(1 + c^2) p^2 - 2 c_2 p + c^2 c_3 + c_2^2}}{c} \nn \\
   &&& \gamma=0 ~~~~\Ra~~~~ q = c p + c_Q \pm\sqrt{(1 + c^2) p^2 - 2 c_2 p - c_3} \nn
 \end{align}
 There are two pairs of intersection points for these two curves, but one pair is complex, depending on the choice of constants:
 \begin{align}
    p_f & = \frac{c_2 \pm \sqrt{(1+c^2)c_3+c_2^2}\;}{c^2+1} ~~~~\mbox{and}~~~~
    q_f = c_Q + \frac{c \big(c_2 \pm \sqrt{(1+c^2) c_3+c_2^2}\; \big)}{c^2+1} ~, \\
    p_f & = \frac{c_2 \pm c\sqrt{-(1+c^2)c_3 - c_2^2}\;}{c^2+1} ~~~~\mbox{and}~~~~
    q_f = c_Q + \frac{c c_2 \mp \sqrt{-(1+c^2)c_3 - c_2^2}\;}{c^2+1} ~.
 \end{align}
 If $\epsilon = -1$ and ~$c_3 < c_2^2/(1 + c^2)$~, the second pair is real, otherwise the first pair.
 The first pair of fixed points are on a line of slope $c$ that passes through $(p, q) = (0,0)$, and the Killing vector field is like that of case 2, but rotated through angle $\tan^{-1} c$.  The second pair are on the circle ~$Y^2 + Z^2 = (1 + c^2) P^2 - 2 c_2 P - c_3 = S^2$~.

 \end{document}